\documentclass[showpacs,amsmath,amssymb,prl,superscriptaddress,floatfix,twocolumn]{revtex4}
\usepackage{graphicx}
\usepackage{dcolumn}
\usepackage{bm}
\usepackage{color}

\begin{document}

\title{Particle focusing in oscillating dissipative billiards} 
\author{Christoph Petri}
\affiliation{Zentrum f\"ur Optische Quantentechnologien, Universit\"at Hamburg, Luruper Chaussee 149, 22761 Hamburg, Germany}
\author{Florian Lenz}
\affiliation{Zentrum f\"ur Optische Quantentechnologien, Universit\"at Hamburg, Luruper Chaussee 149, 22761 Hamburg, Germany}
\author{Fotis K. Diakonos}
\affiliation{Department of Physics, University of Athens, GR-15771 Athens, Greece}
\author{Peter Schmelcher}
\email{Peter.Schmelcher@physnet.uni-hamburg.de} 
\affiliation{Zentrum f\"ur Optische Quantentechnologien, Universit\"at Hamburg, Luruper Chaussee 149, 22761 Hamburg, Germany}
\date{\today}

\begin{abstract}
 We develop and analyze a scheme to achieve both spatial and energetic focusing of an ensemble of neutral particles which is based on an oscillating billiard with frictional forces. The interplay of two competing mechanisms, acceleration due to collisions with the oscillating billiard walls and decceleration  caused by friction, leads to the emergence of attractors in phase space. Their specific properties, i.e. spatial localization and energy spread,  can be controlled and tuned by varying e.g. the frequency of the time-dependent billiard.  
\end{abstract}
\maketitle

Spatial and energetic focusing of neutral particles is an indispensable prerequisite  for many  experimental setups such as atomic \cite{Merimeche:2006} and crossed molecular beams \cite{Lee:1987},  microparticles in a `lab-on-a-chip' \cite{Velev:2003} or in ultracold neutron precision measurements \cite{Ignatovich:1990}. Typical devices used in order to spatially localize particles are lenses, mirrors,  nozzles or collimators. The desired energies are usually achieved by means of velocity selectors or, in the case of neutrons, by moderating media. In this work, we propose a novel scheme for spatial and energetic focusing of an ensemble of neutral particles by using a remarkably simple billiard setup. Necessary ingredients are oscillating walls and frictional forces acting on particles in between collisions with the walls. Their interplay leads to a state of dynamical equilibrium, i.e. periodic and quasi-periodic attractors in phase space, whose  energy and spatial localization can be tuned by varying e.g. the oscillation frequency of the billiard. 

It is well-known that time-dependent billiards \cite{Loskutov:1999,  Carvalho:2006,Lenz:2007, Gelfreich:2008} can be used to study non-equilibrium phenomena like Fermi acceleration \cite{Fermi:1949,Lichtenberg:1992} and anomalous diffusion in momentum space \cite{Lenz:2008, Lenz:2009, Lenz:2009c}. Moreover, dissipation can be introduced to billiard systems either by a restitution coefficient accounting for inelastic collisions \cite{Lieberman:1985, Mehta:1990, Gilet:2009}, or by adding drag forces  between collisions (e.g. Stokes' friction) \cite{Acosta:1990, Leonel:2006, Carvalho:2008}. 

Here, we investigate the particle dynamics in a 2D harmonically  driven elliptical billiard with a frictional force between collisions. There are two competing mechanisms in the presence of driving and dissipation: Firstly, the driving accelerates the particles and secondly, the friction slows them down, i.e. we expect a dynamical equilibrium where both processes balance each other. We will demonstrate that the emergence of attractors focuses the ensemble onto certain parts of phase space.  The position  of the attractors in phase space can be controlled by tuning the parameters of the driving. Consequently, the dissipative driven elliptical billiard can be used as a tool for particle focusing in space and energy without any imaging or focusing devices such as lenses, collimators or monochromators.

The boundary $\mathcal B(t)$ of the driven elliptical billiard is given by $\mathcal{B}(t)=\left \{ \left( a(t) \cos \phi, b(t)\sin\phi \right) | 0\leq \phi <2\pi\right \}$,
where  $a(t)=a_0+C \sin(\omega t)$ and $b(t)=b_0+C \sin(\omega t)$ are assumed to be harmonic functions, where $C>0$, $\omega$ and $a_0,b_0$ are the driving amplitude, the oscillation frequency and  the
equilibrium values of the semi-major and semi-minor axes, respectively.  Due to  friction, particles in-between collisions with the moving boundary do not travel ballistically, but they still move on straight lines: $ \dot{ \bm v}= k \bm v  \Leftrightarrow \bm v(t) = \bm v_0 e^{-kt}$ and the position of a particle (in-between collisions) is simply $\bm x(t)=  \bm x_0 + \frac{\bm v_0}{k}\left (1-e^{-kt}\right)$. We focus on the case (sufficiently small values of $k$), where a stopping of the particles in-between collisions occurs hardly ever.

\begin{figure}
\includegraphics[width=8.6cm]{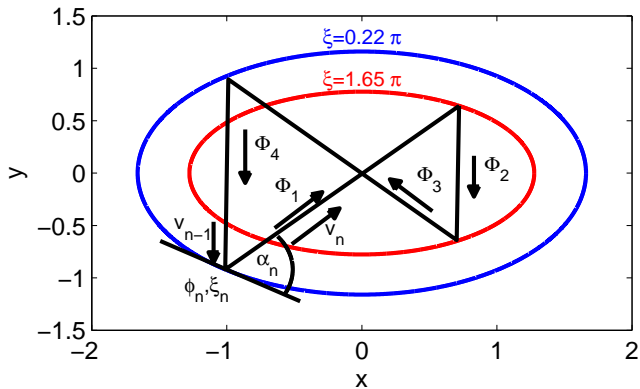}
\caption{(Color online) Boundary of the elliptical billiard in coordinate space at two different phases $\xi$ together with a typical period four attractor ($a_0=1.5, b_0=1, C=0.25, \omega=1$). At the single crossing point, the two fluxes $\Phi_1$ and $\Phi_3$ intersect. \label{fig:fig1}}
\end{figure}

The dynamics  can  be
described by an implicit $4D$ mapping, using the variables
$(\xi_n,\phi_n, \alpha_n, v_n)$, where $v_n=\vert \bm{v}_n \vert$,  $\xi_n = \omega t_n \, (\text{mod} \, 2\pi)$ is
the phase of the boundary oscillation and $\alpha_n$ is the angle between
$\bm{v}_n$ and the tangent of the boundary at the $n$-th collision (see Fig.~\ref{fig:fig1}). The time $t_{n+1}$ of the $n+1$-st collision is determined by the smallest $t_{n+1}>t_n$ that solves the implicit equation
\begin{multline}
 \left ( \frac{a(t_n) \cos \phi_n + v_n^x(1-e^{k \triangle t})/k}{\left [ a(t_{n+1})\right ]^2}\right)^2\\
+\left ( \frac{b(t_n) \sin \phi_n + v_n^y(1-e^{k \triangle t})/k}{\left [b(t_{n+1})\right ]^2}\right)^2-1=0,
\end{multline}
where $\triangle t = t_{n+1}-t_n$ and $\bm v_n= (v_n^x,v_n^y)$. The equations for  $\phi_{n+1}$, $\alpha_{n+1}$ and $v_{n+1}$ are the same as for the elliptical billiard without dissipation, see  Eqs. (9)-(12) of Ref. \cite{Lenz:2009c}.

\begin{figure}
\includegraphics[width=8.6cm]{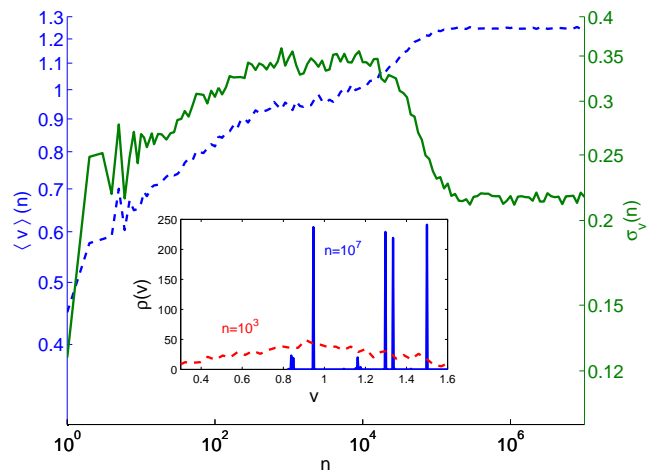}
\caption{(Color online)  $\langle v\rangle (n)$ (dashed) and its standard deviation $\sigma_v (n)$ (solid) (1000 particles with $v_0=0.1$, and $\xi_0, \phi_0,\alpha_0$ distributed uniformly;  $C=0.25$, $a_0=1.5$, $b_0=1$, $k=10^{-3}$). The inset shows the distribution $\rho_n(v)$ at $n=10^3$ (dashed) and at $n=10^7$ (solid). \label{fig:fig2}}
\end{figure}

Without dissipation, the driven elliptical billiard shows FA \cite{Lenz:2008,Lenz:2009, Lenz:2009c}, i.e. the ensemble averaged modulus of the velocity $\langle v \rangle (n)$ grows with increasing number of collisions unboundedly. With friction, this does not hold anymore. The evolution of  $\langle v \rangle (n)$ as a function of the number of collisions $n$ is shown in Fig. \ref{fig:fig2}. Initially ($n<10^3$), $\langle v \rangle (n)$ increases, roughly according to a power law, just like in the elliptical billiard without dissipation. This transient increase is followed by a plateau-like behavior ($10^3<n<10^4$) and a subsequent second increase ($10^4<n<10^5$). Finally ($n>10^5$), the mean velocity $\langle v \rangle (n)$ saturates to a constant value, i.e. there is no FA in the long-term dynamics (this has been checked numerically up to $n\approx10^9$). Naturally, the saturation value depends monotonically on the dissipation parameter $k$ and increases with decreasing $k$.

Intuitively, the absence of FA can be understood in the following way: Between two successive collisions with the oscillating billiard boundary, a particle will slow down by a certain amount $\bigtriangleup v_f$ due to the presence of friction.  For large $v$, the velocity loss $\bigtriangleup v_f$ becomes also large since it is proportional to $v$ itself. On the other hand,  a particle can gain momentum upon a collision with the boundary if the billiard is contracting. However, the velocity gain $\bigtriangleup v_c\leq 2\omega C$ upon  a single collision is independent of $v$. Overall, this means the driven billiard cannot support velocities larger than a typical scale and consequently fast particles will slow down until they are in the velocity range of the attractors.

Effectively, this constitutes high-lying energetically forbidden regions in phase space for particles starting at low velocities. In the beginning of the time-evolution, the particles spread diffusively in phase space, which leads to the transient ($n<10^3$) increase of $\langle v \rangle (n)$. At the same time, the standard deviation $\sigma_v (n)$ of the velocity distribution $\rho_n(v)$ increases as well, see Fig.~\ref{fig:fig2}. The diffusion stops when the ensemble is uniformly spread over the accessible part of phase space, resulting in the plateau-like structure of $\langle v \rangle (n)$ between $10^3$ and $10^4$ collisions. Concurrently, $\sigma_v (n)$ saturates.

The reason for the second increase of $\langle v \rangle (n)$ ($10^4<n<10^5$) is the following: Due to the presence of dissipation,  certain attractors emerge in phase space. These  are either stable limit cycles, corresponding to quasi-periodic motion, or stable spiral points, corresponding to periodic motion.
For the given set of parameters, the velocity of particles on (or very close to) such attractors is roughly between $0.5$ and $1.5$. Step by step, the particles are captured by these attractors, hence the standard deviation $\sigma_v (n)$ decreases in this ($10^4<n<10^5$) regime.  Since the attractors lie rather close in momentum space to the maximum possible energy \cite{Carvalho:2008}, the capturing of the particles onto the attractors is associated with an increase of $\langle v \rangle (n)$. Eventually (at around $n\approx 10^5$), almost all initial conditions are captured by these attractors and consequently $\langle v \rangle (n)$ and $\sigma_v (n)$ approach constant values.

To study the capturing process of the particles onto the attractors further, we investigate the (collision resolved) evolution of the phase space density
\begin{multline}
 \rho_{n_1,n_2}(\phi,\alpha)=\\
\frac{1}{n_2-n_1+1}\sum_{n=n_1}^{n_2}\int_0^{2\pi} d\xi \int_v
dv\rho(\xi,\phi,\alpha,v,n)dv,
\end{multline}
where $\rho(\xi,\phi,\alpha,v,n)=\frac{1}{N_p}
\sum_{i=1}^{N_p}
\delta(\xi-\xi_i^n)\delta(\phi-\phi_i^n)
\delta(\alpha-\alpha_i^n)\delta(v-v_i^n)$ is the phase space density of an ensemble of particles at collision number $n$.

\begin{figure}
\includegraphics[width=8.6cm]{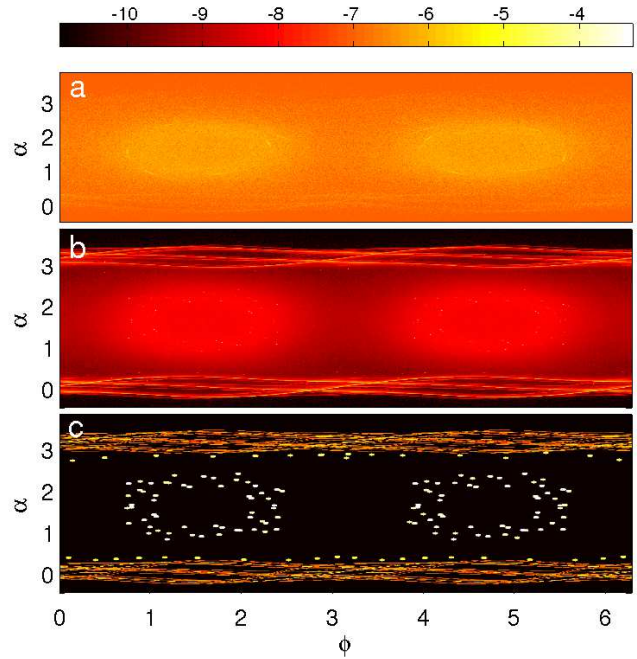}
\caption{\label{fig:fig3}(Color online) Collision resolved evolution of the phase space density (logarithmic colormap) $\rho_{n_1,n_2}(\phi, \alpha)$: a) $10^4<n<2\cdot 10^4$, b) $4\cdot 10^5<n<10^6$ and c) $2\cdot 10^7 <n<10^8$. The ensemble gets focused with increasing number of collisions more and more onto attractors (bright dots and wavy lines in panel c)).}
\end{figure}

In the beginning (Fig.~\ref{fig:fig3}a), the ensemble is distributed almost uniformly over the whole $\phi \times \alpha$ plane. With increasing number of collisions (Fig.~\ref{fig:fig3}b) some particles get focused in the immediate vicinity of the limit-cycle type attractors. These attractors correspond to particles skipping along the elliptical boundary almost tangentially, i.e. they are  whispering gallery orbits and can be observed in Fig.~\ref{fig:fig3}b as wavy lines at $\alpha \approx 0,\pi$. After a very large number of collisions, almost all particles are captured by attractors (Fig.~\ref{fig:fig3}c), i.e. the phase space density is zero outside the regions covered by the attractors.  Now, both types of attractor, periodic orbits (bright dots \footnote{For better visibility, the extension of these bright dots has been magnified artificially}) and limit-cycles are populated.

There co-exist several  periodic attractors with different periods. These  periods are typically between two and thirty (number of collisions with $\mathcal{B}(t)$). Exemplarily, the trajectory of a period four orbit in coordinate space is shown in Fig.~\ref{fig:fig1}, together with the  elliptical boundary at two different phases $\xi$. Since the orbit closes after four collisions, the net energy gained by the four boundary collisions has to exactly match the energy that is dissipated in between these collisions. Furthermore, the orbit has to close after four collisions in the other three variables ($\xi, \phi, \alpha$) as well. Fulfilling all these conditions simultaneously is highly nontrivial, being the reason why there are usually only a few (up to ten) dominant periodic orbit type attractors for a given set of parameters.

The presence of just a few  dominant periodic attractors renders the driven, dissipative elliptical billiard an ideal system in terms of controllability. To remove the vast number of limit-cycle type attractors that undermine this controllability, it is sufficient to introduce a small hyperbolic element (e.g. a small hump) to the boundary of the billiard at $\phi=0$ and $\phi=\pi$. Particles on stable limit-cycles will  hit this hyperbolic element and get scattered back into different parts of phase space. Eventually this process will drive all initial conditions onto the periodic  attractors. Note that we choose $\phi=0,\pi$  for the positions of the hyperbolic elements since generically the periodic attractors do not have collisions with the boundary at these values of $\phi$ (see  Fig.~\ref{fig:fig3}c). The resulting  velocity distribution  $\rho_n(v)$ in the long-term asymptotics ($n>10^6$) is shown in the inset of Fig.~\ref{fig:fig2}. Specifically, there is one dominant  period four attractor (see  Fig.~\ref{fig:fig1}). Each of the four branches of this attractor is associated with a different velocity ($v_i=1.30, 0.95,1.33, 1.50, \, i=1\dots 4$). As a consequence,  $\rho_{n=10^7}(v)$ of the ensemble of particles is not smooth, in contrast to $\rho_{n=10^3}(v)$, but shows distinct peaks at exactly these four values of $v_i$. 

By tuning the parameters of the system, especially the frequency $\omega$, the amplitude $C$ or the friction constant $k$, the position of the attractors in momentum space, and thus $\rho(v)$ can be controlled in a desired manner. Since the frictionless billiards exhibits Fermi acceleration, its dissipative counterpart can possess attractors at arbitrarily large energies. Furthermore, we can populate a certain periodic attractor exclusively by adding additional small humps to the elliptical boundary at positions where particles on other (unwanted) periodic attractors hit the boundary.

With the above described procedure,  particles that are not on the attractor of choice are reinjected  into different parts of phase space until finally all particles are captured by a desired attractor. Thus, ensembles of particles containing few distinct velocity components can be created. By temporarily providing a small opening of the billiard at the collision point of a certain branch of the attractor with the boundary, a monoenergetic spatially focused ensemble can be coupled out of the billiard.

Finally, we investigate the impact of particle-particle interaction on the focusing process onto the periodic attractors. If we assume a short range i.e. contact interaction, there are now two competing processes. Firstly, `background particles' (particles that are not yet on this attractor) get captured onto the attractor with a certain rate $r_c$. Secondly, particles on  the attractor get scattered out of this attractor by two different mechanisms:
\begin{enumerate}
 \item Particles interact with background particles, resulting in a scattering rate $r_b$.
\item  At these crossings points of the attractor in coordinate space, captured particles interact and thus scatter with particles on the same attractor with a certain rate $r_s$. 
\end{enumerate}
This can be modeled by the following system of coupled differential equations for the population $N_a$ of a certain attractor in the long-terms dynamics
\begin{subequations}\label{eq:dgl}
 \begin{align}\label{eq:dgl_a}
\dot N_a &=r_c N_b - r_s N^2_a - r_b N_b N_a\\
\dot N_b &=-r_c N_b + r_s N^2_a + r_b N_b N_a, \label{eq:dgl_b}
\end{align}
\end{subequations}
where $N_b$ is the number of background particles. The capturing rate $r_c$, which of course  depends on the specific attractor,  is determined from our numerical simulations. In the following, we  consider exemplarily the period four  attractor shown in Fig.~\ref{fig:fig1} employing  small humps at $\phi=0,\pi$ to remove the whispering gallery attractors.  To calculate the scattering rate $r_s$, we model the (short range) interaction by a typical scattering cross section $\sigma$. At the crossing point in coordinate space,  there is an incoming flux $\Phi_1$ of particles along branch $1$ and also a flux $\Phi_3$ along branch 3. The number of particles on branch 1 that scatter within a unit time interval at the crossing point is given by $\dot N_1= p_1 \Phi_1 $, where $p_1$ is the probability for a particle on branch 1 to interact with a particle on branch 3. During the time interval in which a particle on branch 1 travels the distance $\sigma$ in order to pass the crossing point,  a particle on branch 3 travels the distance $s=\sigma v_3/v_1$. The scattering probability $p_1$ is then $p_1= (\sigma + s)\Phi_3/v_3$. Since the fluxes obey $\Phi_1=\Phi_3=\Phi$, this yields $\dot N_1= \frac{\sigma (v_1+v_3)}{v_1 v_3}\Phi^2$ and the same expression holds for $\dot N_3$. The flux $\Phi$  can be written as $\Phi=N_a/T_a$, where $T_a$ is the time for one round trip on the attractor. Thus, the change of particles $\dot N_a=-\dot N_1- \dot N_3$ on the attractor due to scattering at the crossing point and the scattering rate $r_s$ (compare with r.h.s. of Eq. \eqref{eq:dgl_a}) are  given by 
\begin{equation}
 \dot N_a= -\frac{2\sigma (v_1 + v_3)}{v_1 v_3 T^2_a} N^2_a \, \Rightarrow \, \, r_s= \frac{2\sigma (v_1 + v_3)}{v_1 v_3 T^2_a}.
\end{equation}
From kinetic theory \cite{Pitaevskii:1981} one obtains for the scattering rate $r_b$ between background particles and particles on an attractor
\begin{equation}
 r_b= \frac{\sqrt{2}\sigma \langle v_b \rangle}{A_{el}},
\end{equation}
where $\langle v_b \rangle$ is the mean  velocity of the background particles and $A_{el}= \frac{\pi}{T}\int_{0}^{2\pi} a(t)b(t)dt=\pi(a_0 b_0+C^2/2)$ is the mean area of the elliptical billiard.  

The system of differential equations~\eqref{eq:dgl} possesses stable solutions that approach asymptotically a constant value. Specifically, we obtain numerically $r_c=3\cdot 10^{-5}$ and for $\sigma=2\cdot 10^{-5}$ the scattering rates are  $r_s= 0.08\sigma$ and $r_b=0.3\sigma$ ($\langle v_b\rangle=1$) yielding a high population of $N_a=94.3\%$.  As long as  $\sigma$ is smaller than $10^{-4}a_0$, the dynamics, especially the focusing of the ensemble onto attractors is not disturbed significantly, implying the possibility of a controlled preparation of the ensemble. 

We demonstrated that combining dissipation and driving in a billiard leads to the emergence of periodic attractors in phase space onto which the originally uniformly distributed ensemble gathers. Our remarkably simple setup therefore allows spatial and energetic focusing of particles and might represent an experimental alternative to  devices such as  lenses, collimators and monochromators. In principle, the billiard setup  could be of macroscopic or mesoscopic character. One possible realization is an atom-optical experiment, since there, the geometry of the billiard  (a rapidly scanning blue-detuned laser) can be  varied arbitrarily in time using acousto-optical modulators and dissipation can be introduced via optical molasses \cite{Milner:2001}. 

P. S. and C. P. acknowledge support from the German Research
Foundation (DFG). Fruitful discussions with S. Hofferberth are gratefully acknowledged by P. S.

\end{document}